# Optimal Preprocessing for Joint Detection and Classification of Wireless Communication Signals in Congested Spectrum Using Computer Vision Methods

Xiwen Kang
Intelligent Fusion Technology, Inc.
20410 Century Blvd.
Germantown, MD 20874

Hua-mei Chen
Intelligent Fusion Technology, Inc.
20410 Century Blvd.
Germantown, MD 20874

Genshe Chen
Intelligent Fusion Technology, Inc.
20410 Century Blvd.
Germantown, MD 20874

Kuo-Chu Chang
Department of SEOR
George Mason University
4400 Univ. Dr.
Fairfax, VA 22030

Thomas M. Clemons
Department of SEOR
George Mason University
4400 Univ. Dr.
Fairfax, VA 22030

*Abstract*— The joint detection and classification of RF signals has been a critical problem in the field of wideband RF spectrum sensing. Recent advancements in deep learning models have revolutionized this field, remarkably through the application of state-of-the-art computer vision algorithms such as YOLO (You Only Look Once) and DETR (Detection Transformer) to the spectrogram images. Building on our previous work that pioneered the use of YOLOv8 for signal detection and classification in congested spectrum environments, this follow-up paper focuses on optimizing the preprocessing stage to enhance the performance of these computer vision models. Specifically, we investigated the generation of training spectrograms via the classical Short-Time Fourier Transform (STFT) approach, examining four classical STFT parameters: FFT size, window type, window length, and overlapping ratio. Our study aims to maximize the mean average precision (mAP) scores of YOLOv10 models in detecting and classifying various digital modulation signals within a congested spectrum environment. Firstly, our results reveal that additional zero padding in FFT does not enhance detection and classification accuracy and introduces unnecessary computational cost. Secondly, our results indicated that there exists an optimal window size that balances the trade-offs between and the time and frequency resolution, with performance losses of approximately 10% and 30% if the window size is four or eight times off from the optimal. Thirdly, regarding the choice of window functions, the Hamming window yields optimal performance, with non-optimal windows resulting in up to a 10% accuracy loss. Finally, we found a 10% accuracy score performance gap between using 10% and 90% overlap. These findings highlight the potential for significant performance improvements through optimized spectrogram parameters when applying computer vision models to the problem of wideband RF spectrum sensing.

I. INTRODUCTION

The field of wideband RF spectrum sensing has been revolutionized by recent breakthroughs in deep learning. Various deep learning frameworks, including Convolutional Neural Networks (CNNs), Recurrent Neural Networks (RNNs), Long Short-Term Memory networks (LSTMs), and transformer networks, have been leveraged to extract critical RF signal parameters like modulation schemes, bandwidth/symbol rates, and carrier frequency offsets [1] – [5]. Among these various methods, one particularly innovative and promising approach is the application of state-of-the-art (SoTA) computer vision models to the RF spectrograms for the challenging task of joint detection and classification of RF signals in dynamic spectrum environments. In [6] and [7], SoTA computer vision models such as YOLOv5 (You Only Look Once) and DETR (Detection Transformer) were applied to the Sig53 and Wideband-Sig53 (WBSig53) synthetic RF datasets to detect the presence, time, frequency, and modulation family of all signals present in the input data. The WBSig53 datasets included 53 classes of modulation schemes and hardware-related impairments such as I/Q imbalance and RF roll-off. The SPREAD dataset in [8] and WBR-DE dataset in [9] are some of the latest datasets that included communication and radar signals in congested environments. Similar to these efforts, our previous work in [10] diversified and enhanced the complexity of the synthetic RF datasets by introducing the random occurrence of time and frequency domain overlapped signals, which better reflect the congested nature of modern RF spectrum environments observed both in civilian and military settings. By applying the SoTA YOLOv8 model to our proposed datasets, we have demonstrated the remarkable potential of advanced CV models in addressing spectrum sensing challenges in congested and dynamic RF environments involving both communication and radar signals. While these previous efforts have shown great promise in jointly detecting and classifying RF signals, the preprocessing stage, specifically the generation of spectrograms via the classical Short-Time Fourier Transform (STFT) method [11], remains a critical area for optimization.

This paper focuses on optimizing the STFT parameters to enhance the performance of SoTA CV models like YOLOv10 [12] in joint detection and classification tasks. The standard STFT for spectrogram generation involves several critical parameters: FFT size, window type, window length, and overlapping ratio. The importance of selecting appropriate STFT parameters is highlighted in various studies across different domains. For instance, in [13], while addressing the audio source separation problem, the author remarked that any machine learning approach applied to spectrogram representations and source separation problems would likely benefit from at least careful consideration of the time-frequency trade-off. The author advocated for tailored STFT window sizing based on the characteristics of the source signals, and argued that the spectrogram time and frequency resolution to some extent defines the degree to which the overlap of signal energy occurs. We found the author's remarks extremely insightful and directly applicable to wideband RF signal processing in congested spectrum environments, where overlapping signals in both time and frequency domains necessitate careful parameter selection.

Furthermore, in [14], the authors proposed an adaptive approach to select the appropriate window length for STFT and get the optimal SNR with the right time-frequency resolution according to the signal characteristic under a fixed sampling rate. Similarly, in [15], the authors proposed an STFT processor that provides a 0/25/50/75% variable overlap ratio to minimize data loss depending on the type of window used and 16/64/256/1024-point variable window lengths to support various time–frequency resolutions. All these recent studies have underscored the critical need to thoroughly examine how the selection of STFT parameters affects the performance of CV models in detecting and classifying RF signals using spectrograms.

Our contributions in this paper are threefold: Firstly, through extensive experiments on diverse wireless communication datasets with various types of modulations in a dynamic and congested RF environment, we thoroughly investigated the effects of classical STFT parameters on the performance of state-of-the-art computer vision methods like YOLOv10 when applied to the joint detection and classification of wireless communication signals. Secondly, we provided a quantitative characterization of the performance gap between optimized and non-optimized STFT parameters, illustrating which parameters are more crucial and sensitive in affecting the mean average precision (mAP) scores of the CV model. Finally, our results lay the foundation for and invite further research on several potential follow-up topics, such as establishing a quantitative relation between the optimal window size and the characteristics of source RF datasets (e.g., the minimum signal duration and minimum signal bandwidth in the source datasets).

The remaining sections of the paper is organized as follows: Section II revisits our previous work to introduce our synthetic RF signal generator, capable of simulating a dynamic, congested, and contested RF spectrum environment with multiple signal types. Section III explains the preparation of the training datasets by detailing how different parameters are applied in the STFT generation stage and comparing the resulting spectrogram images. Section IV evaluates the performance and results of the different parameter choices. Finally, Section V presents our conclusions and explores potential directions for future research.

## II. RF SIGNAL EMULATOR

### A. Signal Emulator Capabilities

Previously in [10], we introduced our RF signal emulator capable of generating of 26 different modulations and signal types commonly used in communication, radar, and navigation applications. Additionally, our emulator supports a dynamic mode where the key signal parameters, such as bandwidth/symbol rate, carrier frequency, signal power, and transmission duration, can be set to vary continuously to emulate a dynamic spectrum environment.

### B. Example Output Waveform and Spectrogram

Fig. 1 shows the time domain waveforms and frequency domain STFT spectrograms of six different digital communication signals generated using our emulator. For easy interpretation of the time-domain waveforms, we have set the adjustable carrier frequency to zero (i.e., at the baseband).

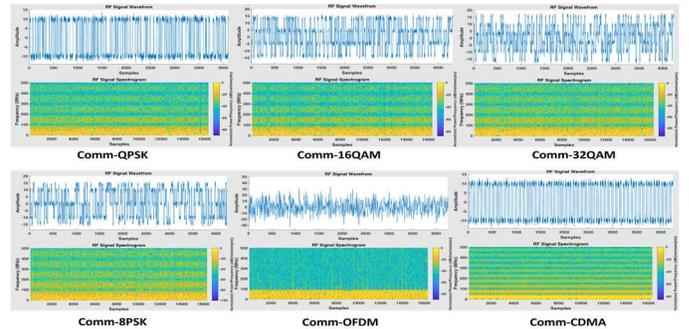

Fig. 1. Example output of the RF simulator (showing the baseband I-component only).

In addition to showing each individually generated signal, the more interesting feature of our RF emulator is its capability to emulate the combined signal observed by a receiver in a congested spectrum environment. As an example, Fig. 2 demonstrated a scenario where there are eight coexisting RF emitters simultaneously transmitting different digital modulation waveforms within the 500 MHz band. To apply the computer vision-based methods for joint signal detection and classification tasks, each instance of signals in the training spectrograms are enclosed by a bounding box with height and width determined based on the signal's effective bandwidth and active duration. The bounding box's vertical position is determined by the carrier frequency and the horizontal position is determined by the transmission start and end times.

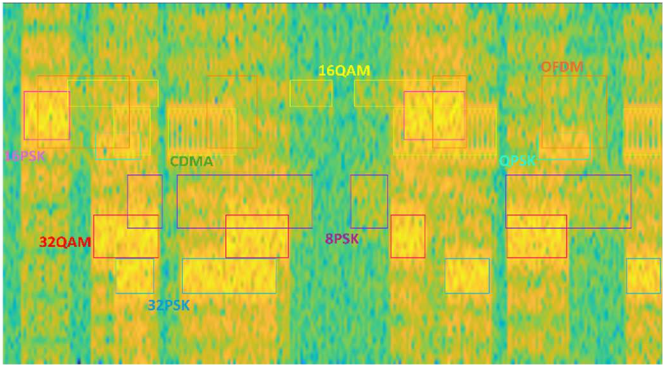

Fig. 2. Annotated spectrogram with bounding boxes showing different digital modulation signals coexisting with each other.

### III. EXPERIMENT SETUP AND DATASET PREPARATION

In this section, we will describe the simulation setup and the preparation of the training datasets for each STFT parameter under study.

#### A. Simulation Setup

Our simulation setup is similar to what was used in section III.A of [10]. We again simulated a dynamic spectrum environment consisting of 8 classes digital modulation signals with varying signal parameters. The exact range for each parameter is summarized in TABLE I. below. The only difference from [10] is that we slightly increased the spectrum congestion level by expanding the maximum allowable signal bandwidth from 60 MHz to 100 MHz.

TABLE I. SIMULATION SETUP FOR CONGESTED DIGITAL MODULATION SIGNALS

| | |
|---|---|
| Modulation Schemes | QPSK, 8PSK, 16PSK, 32PSK, 16QAM, 32QAM, CDMA-QPSK, OFDM-QPSK |
| Number of Samples (per Spectrogram) | 4096 samples/timeslot * 4 timeslots = 16384 samples |
| Number of Spectrograms Generated | 4000 |
| Carrier Frequency (MHz) | 100 MHz < fc < 400 MHz |
| Single-sided Bandwidth/Symbol Rate (MHz) | 20 MHz < BW < 100 MHz |
| Transmission Duration (% of timeslot duration) | 20% < Dt < 100% |
| SNR (dB) | 0 dB < SNR < 25 dB |

#### B. Dataset Preparation Process

Considering the computational challenges associated with jointly optimizing all four STFT parameters, we adopted a sequential optimization approach. Initially, we conducted a joint coarse grid search on window length and FFT size using an initial batch of datasets. After determining that FFT size had no significant effect, we proceeded with a fine search to identify the optimal window length. Subsequently, we optimized the window type based on the previously determined window length, and finally, we optimized the overlapping factor using the optimized window length and type.

*1) Initial datasets for coarse grid search of optimal window length and FFT size*

Our first batch of datasets are generated to help conduct an initial coarse grid search of the optimal window length and FFT size. Since the time-domain waveform has the length $N = 16384$ samples, we investigated window sizes $W = \frac{N}{4}, \frac{N}{16}, \frac{N}{64}, \frac{N}{256}$ and FFT sizes (with zero-padding) $F = 4W, 16W, 64W, 256W$. The complete list of parameter pairs is shown in TABLE II. TABLE II. below.

TABLE II. INITIAL COARSE GRID SEARCH FOR OPTIMAL WINDOW LENGTH AND FFT SIZE

| W64F256 | W64F1024 | W64F4096 | W64F16384 |
|---|---|---|---|
| W256F1024 | W256F4096 | W256F16384 | W256F65536 |
| W1024F4096 | W1024F16384 | W1024F65536 | W1024F262144 |
| W4096F16384 | W4096F65536 | W4096F262144 | W4096F1048576 |

Fig. 3 below shows the example output spectrograms obtained using the STFT parameter pairs listed above. The default window type is the Hamming window and 50% overlap ratio is used. It is clear that the window length has a much greater effect in shaping the final spectrograms.

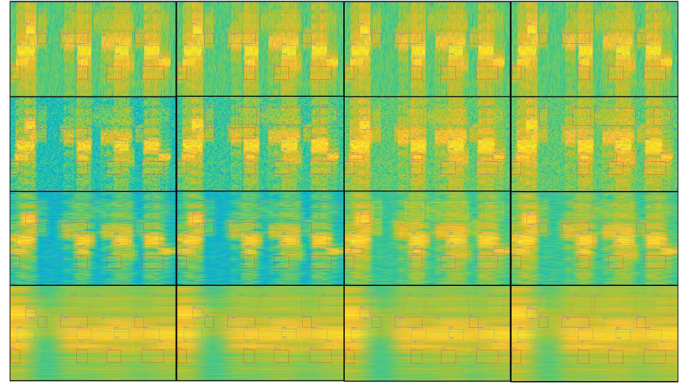

Fig. 3. Example output spectrograms for the initial coarse grid search.

Once the STFT process is complete, the 4,000 generated spectrograms (for each STFT parameter setting) are divided into training, validation and test sets, each containing 2,800, 800, and 400 images, respectively.

*2) Additional datasets for fine search of optimal window length*

Following the initial grid search and additional experiments (see details in Section IV), we have concluded that additional zero padding in FFT does not enhance detection and classification accuracy and introduces unnecessary computational cost. Therefore, we have created the second batch of datasets (with no zero-padding) for fine search of the optimal window length. The complete list of parameter pairs is shown in TABLE III. below.

TABLE III. FINE SEARCH FOR OPTIMAL WINDOW LENGTH

| W8F8 | W16F16 | W32F32 | W64F64 |
|---|---|---|---|
| W128F128 | W256F256 | W1024F1024 | W4096F4096 |

Fig. 4 below shows the example output spectrograms obtained using the STFT parameter pairs listed above. It is clear that there exists an optimal window length with the best time-frequency resolution trade-off.

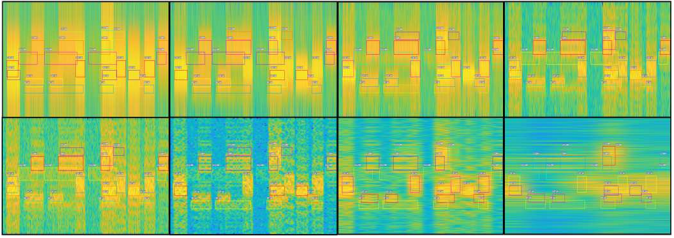

Fig. 4. Example output spectrograms for the fine search of optimal window length.

*3) Additional datasets for fine search of optimal window type*

With the optimal FFT size and window length determined, our third batch of datasets are generated using various types of windowing function. The list of window types investigated is shown in TABLE IV. below.

TABLE IV. FINE SEARCH FOR OPTIMAL WINDOW TYPE

| Hann window | Gaussian window | Hamming window |
|---|---|---|
| Blackman window | Rectangular window | Bohman window |
| Tapered cosine window | Flat Top window | Nuttall's Blackman-Harris window |

Fig. 5 below shows the example output spectrograms obtained using window functions listed above. Some window functions, such as the flat top window, yield more blurry spectrograms compared with the other window selections.

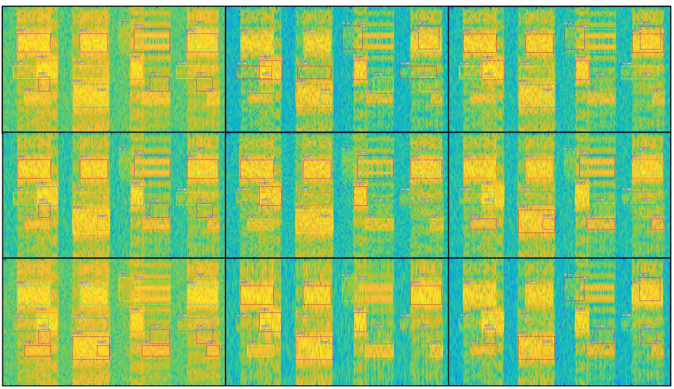

Fig. 5. Example output spectrograms for the fine search of optimal window type.

*4) Additional datasets for fine search of optimal overlap factor*

Our final batch of datasets are generated using various choices of overlap factors. Although 50% overlap factor is typically recommended, we still want to know the amount of potential performance improvement from using higher overlap ratios. The list of overlap factors investigated is shown in TABLE V. below.

TABLE V. FINE SEARCH FOR OPTIMAL OVERLAP RATIO

| 10% | 20% | 30% |
|---|---|---|
| 40% | 50% | 60% |
| 70% | 80% | 90% |

Fig. 6 below shows the example output spectrograms obtained using various overlap factors listed above.

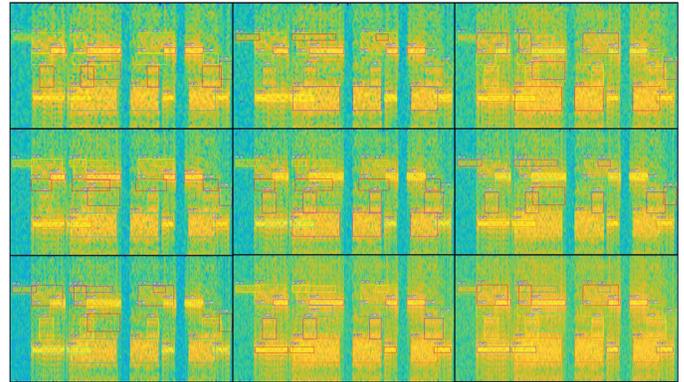

Fig. 6. Example output spectrograms for the fine search of optimal overlap factor.

## IV. PERFORMANCE EVALUATION

In this section, we present and discuss the results obtained from each batch of our parameter optimization datasets. For each dataset, we fine-tuned the Ultralytics pre-trained YOLOv10s model on a single NVIDIA Tesla T4 GPU, utilizing a training set of 2,800 spectrogram images. The model was fine-tuned for 50 epochs, employing an AdamW optimizer with a batch size of 16 and a learning rate of 8.33e-4.

### A. Coarse grid search of optimal window length and FFT size

The mAP performance scores for the coarse grid search of optimal window length and FFT size are summarized in TABLE VI. below. Based on the initial coarse optimization results, we made two observations. First, there is no noticeable detection and classification performance gain with the usage of zero-padding and a greater number of total FFT points. This result in fact could appear counter-intuitive. Although it is a well-known fact that the usage of zero-padding does not increase the "true" resolution or increase the amount of information present in the spectrograms, it does increase the interpreted "visual resolution" of the spectrograms. Similar to oversampling in the time domain, which leads to a smoother appearance of a linearly interpolated waveform, increasing the FFT size would also result in a smoother appearance of the spectrogram image. Our results suggested that the SoTA CV models are able to fully extract the available amount of information from the spectrograms without the need for additional zero-padding, and using an FFT size greater than the original window size may lead to wasted computational resources.

TABLE VI. COARSE GRID SEARCH RESULTS (MAP50 / MAP50-95)

| W \ F | 4W | 16W | 64W | 256W |
|---|---|---|---|---|
| N/256 | 0.841/0.690 | 0.831/0.682 | 0.823/0.677 | 0.835/0.684 |
| N/64 | 0.815/0.672 | 0.822/0.678 | 0.815/0.673 | 0.826/0.685 |
| N/16 | 0.644/0.481 | 0.643/0.478 | 0.636/0.472 | 0.635/0.472 |
| N/4 | 0.173/0.081 | 0.171/0.083 | 0.176/0.086 | 0.178/0.084 |

Our second observation is that there exists an optimal window size that yields the best detection and classification peformance by balancing the trade-offs between and the time and frequency resolution. With the reasonable assumption that there is a monotonic trend near the optimal point (i.e., performance improves steadily as it nears the optimum and declines steadily after surpassing it), we can observe that the optimal widnow size must be less than N/64 =256.

*B. Fine search of optimal window length*

Fig. 7 illustrated the results we obtained for our second batch of datasets consisting of spectrograms generated using various choices of window size. Out of all the window sizes investigated, N/128=128 yields the optimal performance with mAP50 score of 0.858 and mAP50-95 score of 0.706. Here we make the observation that the optimal window size (out of all the window sizes investigated) is also equal to the square root of the total number of samples, i.e.:

$$w_{opt} = \sqrt{N_{sample}} \quad (1)$$

As pointed out in our earlier review of relevant literature, the exact optimal choice of STFT window size will always depend on the characteristics of the source datasets. For our particular synthetic datasets, we have the ratio between the minimum signal bandwidth and the source I/Q data sampling bandwidth given by:

$$R_f = \frac{BW_{min}}{BW_{sample}} = \frac{40}{500} = 0.08 \quad (2)$$

And the ratio between the minimum signal active duration to the total sample length is given by:

$$R_t = \frac{N_{min}}{N_{sample}} = D_{t,min} \cdot \frac{1}{\# \text{ of timeslot}} = 0.05 \quad (3)$$

For future research, it may be useful to characterize the source RF datasets with the following definition of time-frequency skewness:

$$\mu_{tf} = \frac{R_t}{R_f} \quad (4)$$

As we can observe, for our relatively balanced (i.e., $R_f$ and $R_t$ have the same order of magnitude) source datasets, $w = \sqrt{N_{sample}}$ offers balanced resolutions in both time and frequency domains. Nevertheless, a non-unity skewness of 0.625 still results in a slightly lower window size being more favored. This is illustrated by the fact that we have mAP50-95 score of 0.69 at $w = 64$ and the score of 0.687 at $w = 256$. We will leave it as part of our future work to derive a quantitative relation between the optimal window size and the characteristics (e.g., minimum signal BW, sampling bandwidth, minimum signal active duration, total duration of each spectrogram) of the source datasets.

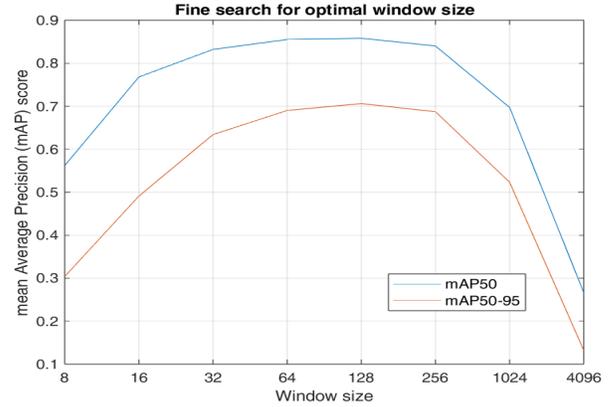

Fig. 7. Detection and classification performance for various choices of window length.

Additionaly, results in Fig. 7 also highlighted the sensitivty of the CV model's detection and classification performance to the choice of STFT window length. For window lengths that were four or eight times off from the optimal value, we observed performance degradations of approximately 10% and 30%, respectively.

*C. Fine search of optimal window type*

Following our investigation of the optimal window size, we studied the performance yielded by various choices of windowing function. TABLE VII. below summarizes the mAP performance scores achieved using a variety of commonly used STFT window types.

TABLE VII. RESULTS FOR VARIOUS WINDOW TYPES (MAP50 / MAP50-95)

| Window Type | Hann | Gaussian | Hamming |
|---|---|---|---|
| mAP Scores | 0.854/0.699 | 0.848/0.695 | 0.858/0.706 |
| Window Type | Blackman | Rectangular | Bohman |
| mAP Scores | 0.844/0.685 | 0.83/0.672 | 0.836/0.68 |
| Window Type | Tapered cosine | Flat Top | Nuttall |
| mAP Scores | 0.838/0.686 | 0.823/0.644 | 0.828/0.668 |

Among all the windowing functions investigated, we found that the STFT spectrograms generated using the Hamming window, which minimizes the nearest side lobe and is also one of the most commonly used windows, yielded the optimal detection and classification performance with mAP50 score of 0.858 and mAP50-95 score of 0.706. In terms of the performance gap, our results showed that using a nonoptimal window, such as the flat top window, could lead to an accuracy loss of up to 10%.

*D. Fine search of optimal overlap factor*

Our final study focused on the effects of STFT overlap factor on the detection and classification performance of the CV-based methods. Fig. 8 illustrated the results we obtained from various choices overlap factor ranging from 10% to 90%. As we would expect, higher overlap factors indeed lead to slightly better mAP performance scores at the expense of increased computational demand. Since we have observed less than a 10% difference in accuracy between using 10% and 90% overlap factors, it is therefore reasonable to adopt the commonly chosen 50% overlap factor in real-world applications.

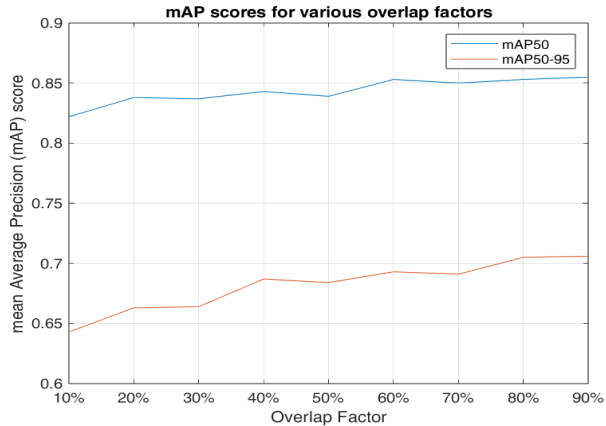

Fig. 8. Detection and classification performance for various choices of STFT overlap factor.

## V. CONCLUSION AND FUTURE WORK

In this work, we have demonstrated how the selection of the spectrogram pre-processing parameters can affect the final detection and classification accuracy performance of the computer-vison based spectrum sensing methods. In addition to investigating the optimal window length, window type, FFT size and overlap factor that maximize the mAP scores for our synthetic RF datasets, we have evaluated and highlighted the potential accuracy performance loss resulted from using spectrograms generated with non-optimal choices of STFT parameters.

Regarding the future work, we will continue to expand and diversify our synthetic RF datasets to cover a wide range of time-frequency skewness levels as defined in Subsection B of Section IV. The ultimate goal is to derive a relatively generalized quantitative relationship between the characteristics of the source datasets and the optimal choice of window length.

## VI. SOFTWARE AND DATA

All datasets (spectrograms and bounding box labels) mentioned in Section III and used for obtaining our results have been shared publicly on Roboflow. The links to the Roboflow datasets can be found at: https://github.com/xwkang2019/Optimal_PreProcessing_for_CV_based_spectrum_sensing/